\documentclass[aps,prc,groupedaddress,showpacs,twocolumn]{revtex4}

\usepackage{graphicx}

\begin{document}

\title{The effect of ``pre-formed" hadron potentials on the dynamics of heavy ion collisions and the HBT puzzle}

\author {Qingfeng Li,$\, ^{1}$\footnote{E-mail address: liqf@fias.uni-frankfurt.de}
Marcus Bleicher,$\, ^{2}$ and Horst St\"{o}cker$\, ^{1,2,3}$}
\address{
1) Frankfurt Institute for Advanced Studies (FIAS), Johann Wolfgang Goethe-Universit\"{a}t, Max-von-Laue-Str.\ 1, D-60438 Frankfurt am Main, Germany\\
2) Institut f\"{u}r Theoretische Physik, Johann Wolfgang Goethe-Universit\"{a}t, Max-von-Laue-Str.\ 1, D-60438 Frankfurt am Main, Germany\\
3) Gesellschaft f\"ur Schwerionenforschung, Darmstadt (GSI), Germany \\
 }


\begin{abstract}
The nuclear stopping, the elliptic flow, and the HBT interferometry
are calculated by the UrQMD transport model, in which potentials for
``pre-formed" particles (string fragments) from color fluxtube
fragmentation as well as for confined particles are considered. This
description provides stronger pressure at the early stage and
describes these observables better than the default cascade mode
(where the ``pre-formed" particles from string fragmentation are
treated to be free-streaming). It should be stressed that the
inclusion of potential interactions pushes down the calculated HBT
radius $R_O$ and pulls up the $R_S$ so that the HBT time-related
puzzle disappears throughout the energies from AGS, SPS, to RHIC.
\end{abstract}


 \maketitle

One of the most important discoveries from SPS and RHIC experiments
in the beginning of this century is that a new form of matter is
produced in high energy ion collisions that seems to behave as a
strongly interacting quark gluon plasma (sQGP)
\cite{Lee:2005gw,Gyulassy:2004zy,Shuryak:2004cy,Satarov:2005mv}.
Furthermore, at RHIC energies, the sQGP seems to be a nearly
``perfect liquid" with small viscosity
\cite{Janik:2005zt,Hirano:2005wx}. Although hydrodynamical studies
are a useful tool for the theoretical investigations at RHIC
energies, hydrodynamics fails at lower energies. Therefore, a
relativistic dynamic transport approach, in which an entire
equilibrium is not pre-assumed, is necessary to explore the
excitation functions of various observables. In addition, the
transport model has the advantage of observing the whole phase-space
evolution of all particles involved in a microscopic fashion.

In order to detect this new matter - quark gluon plasma (QGP) - many
theoretical suggested observables have been discussed and argued
frequently. Among these, the Hanbury-Brown-Twiss interferometry
(HBT) or Femtoscopy technique has been used widely to extract the
spatio-temporal information of the particle freeze-out source. It
would be a vital discovery if there is an nontrivial transition in
the spatio-temporal characteristics of the source, when going from
low to high beam energies \cite{Rischke:1996em}. Experimentally, the
HBT parameters have been scanned thoroughly over the energies from
SIS, AGS, SPS, up to RHIC, unfortunately, no obvious discontinuities
do appear \cite{Lisa:2005dd}.

However, this ``null" result does not mean that the HBT technique
comes to an end. Non-Gaussian effects might shadow the possible
energy dependence of the HBT parameters (``{\it E-puzzle}"), and
this topic has been quickly improving in the recent years
\cite{Lin:2002gc,Chung:2007yq,Danielewicz:2006hi}. Secondly, even
with a Gaussian parametrization, we found that the HBT time-related
puzzle (``{\it t-puzzle}") is present at almost all energies from
AGS to RHIC \cite{Li:2007im}. In order to understand the origins of
these ``puzzles", it is necessary to dig deeper into the dynamics of
the heavy ion collisions.

The hydrodynamic model, in which various kinds of equations of state
(EoS) with latent heats are considered, successfully explained the
elliptic flow $v_2$ at transverse momentum $p_t<2$GeV$/c$ at RHIC
energies \cite{Huovinen:2003fa,Kolb:2003dz,Huovinen:2005gy}.
However, hydrodynamics failed to explain the experimental $R_O/R_S$
data (\cite{Lisa:2005dd} and the references therein), i.e. the ratio
between the HBT radii in outward and sideward directions, where the
``t-puzzle" initially occurred. But recently, it has been realized
that a direct (fully ``apple-to-apple") comparison between hydro
calculations and data is required \cite{Frodermann:2006sp}.

In the transport models, the effective EoS is softened by the
existence of resonances, but the re-scattering between particles
will reinforce to some extent the stiffness of the dense matter. At
SPS and RHIC energies, the color-string excitation and fragmentation
is further considered for particle production and subsequent
evolution in the ultra-relativistic/relativistic quantum molecular
dynamics (UrQMD/RQMD) hadron-string transport models. The formation
time of the hadrons from string fragmentation is determined by a
``yo-yo" mode \cite{Bass98,Bleicher99}. During the formation time,
the ``pre-formed" particles (string fragments that will be projected
onto hadron states later on) are usually treated to be
free-streaming, while reduced cross sections are only included for
leading hadrons. The idea of ``pre-formed" hadrons is a well-known
concept in the description of deep inelastic scattering data (DIS)
\cite{Czyzewski:1989ur,Accardi:2002tv}. It has also been found that
the ``pre-formed" hadrons from the color fluxtube fragmentation
behave already very hadronlike, before the hadron formation is fully
completed \cite{Bialas:1986cf}. In previous calculations
\cite{Bass98,Bleicher99,Bratkovskaya:2004kv}, the interaction of the
newly created ``pre-formed" particles from the string fragmentation
is usually not taken into account for simplicity. The facts have
shown that the effective EoS at the early stage in the past model
simulations is too soft and the early pressure is inadequate to
match the calculated elliptic flow to the experimentally observed
flows at RHIC energies \cite{Zhu:2005qa,Bleicher:2000sx}. In order
to make up for this discrepancy in the early-stage EoS, it is
essential to modify the dynamics of the early stage with/without new
degrees of freedom, such as done in a multi-phase transport (AMPT)
\cite{Lin:2002gc}, the hadronic rescattering (HRM)
\cite{Humanic:2005ye}, the quark molecular dynamics (qMD)
\cite{Hofmann:1999jx}, and the parton hadron string dynamics (PHSD)
\cite{Arsene:2006vf} model.

Besides the new matter which is dominant at the early stage, the
potentials between confined hadrons which play a role during almost
all time of the collision process deserve attention as well. In the
previously low energy UrQMD calculations, the Hamiltonian $H$ served
for the EoS consists of the kinetic energy $T$ and the effective
two-body interaction potential energy $U$. The potential energy $U$
is indispensable, which includes the two-body and three-body (which
can be approximately written in the form of two-body interaction)
Skyrme- (also called as the density dependent terms), Yukawa-, and
Coulomb-terms \cite{Bass98,Bleicher99}. Recently, in order to be
more successfully applied in the intermediate energy region ($E_b
\lesssim 2$A GeV), more terms are considered
\cite{Li:2003zg,Li:2005gf}, those are, the density-dependent
symmetry potential (essential for isospin-asymmetric reactions at
intermediate and low energies) and the momentum-dependent term. At
higher energies, i.e., AGS/FAIR and SPS, the Yukawa- and symmetry-
potentials of baryons becomes negligible, while the Skyrme- and the
momentum-dependent part of potentials still influence the whole
dynamic process of HICs though weakly. In Ref.~\cite{Isse:2005nk},
the effects of the mean field with momentum dependence on collective
flows from HICs at 2-158A GeV energies were studied by a Jet AA
Microscopic Transportation Model (JAM) and it has been found that
the momentum dependence in the nuclear mean field is important for
the understanding of the proton collective flows at AGS and even at
SPS energies. Further, by adopting the same soft EoS with momentum
dependence (SM-EoS) on confined baryons, we also found that the
calculated $R_O/R_S$ ratio from HICs at AGS energies can be vastly
improved when compared to the measured data
\cite{Li:2007im,lqf20063}.

In this work, we will follow this idea and investigate the HBT radii
at higher beam energies, i.e. SPS and RHIC. Furthermore, the other
observables, such as the rapidity distributions of net protons
($p-\bar{p}$) and pions, and the elliptic flows of charged hadrons,
are checked at the same time. As an initial attempt, we consider
potentials for both formed and ``pre-formed" particles: 1), for
formed baryons, the SM-EoS \cite{Isse:2005nk} is chosen. The Coulomb
potential is also considered, as in the low-energy calculations
\cite{Li:2005gf,lqf20063}. 2), for formed mesons, only the Coulomb
potential is optional in calculations, while other potentials are
not considered as in previous calculations at low energies
\cite{Li:2005gf,lqf20063}. 3), for ``pre-formed" particles from
string fragmentation, the similar density dependent (Skyrme-like)
terms \footnote{$U(\rho_h/\rho_0)=\alpha (\rho_h/\rho_0)+\beta
(\rho_h/\rho_0)^\gamma$, where $\rho_h$ is the hadronic density,
$\rho_0=0.16 fm^{-3}$ is the normal nuclear density. $\alpha$,
$\beta$, and $\gamma$ are parameters, in this work for the SM-EoS,
they are, $-268$ MeV, $345$ MeV, and $7/6$, respectively.} as the
formed baryons are used, but without the Yukawa, the Coulomb, and
the momentum dependent terms. 4), the ``pre-formed" mesons act like
``pre-formed" baryons but with a reduction factor ($2/3$) due to the
quark-number difference. Furthermore, 5), there is no potential
interaction between ``pre-formed" baryons and formed baryons so far.
Correspondingly, the hadronic density is calculated by
$\rho_h=\sum_{j\neq i}c_i c_j\rho_{ij}$, where $c_{i,j}=1$ for
Baryons, $2/3$ for ``pre-formed" mesons, and $0$ for formed mesons.
$\rho_{ij}$ is a Gaussian in coordinate space. As stated in
\cite{Isse:2005nk,lqf20063}, the relativistic effect (Lorentz
transformation) on the relative distance and relative momentum
between two particles $i$ and $j$ has been considered.

\begin{figure}
\includegraphics[angle=0,width=0.4\textwidth]{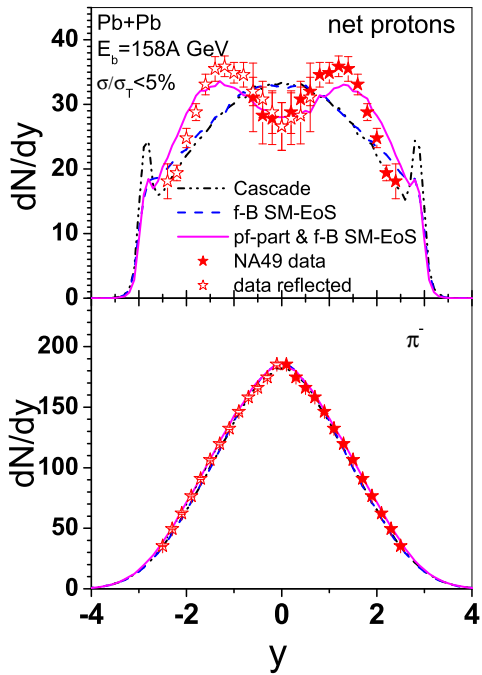}
\caption{Rapidity distributions of net protons (top) and negatively
charged pions (bottom) from central ($<5\%$ of total cross section)
Pb+Pb collisions at $E_b=158$A GeV. The experimental data (solid
stars) are taken from \cite{Afanasiev:2002mx,Appelshauser:1998yb},
while the open stars represent the data reflected at $y=0$ in the
center-of-mass system. The dash-dot-dotted lines are the
calculations with cascade mode, the dashed lines represent the
calculations with potentials for confined baryons (SM-EoS), while in
the solid lines, the density-dependent potentials for ``pre-formed"
particles are further taken into account. } \label{fig1}
\end{figure}

To set the stage, Fig.\ \ref{fig1} shows the rapidity dependence of
the multiplicities of net protons (top) and negatively charged pions
(bottom) for central ($<5\%$ of total cross section $\sigma_T$)
Pb+Pb collisions at $E_b=158$A GeV. The experimental data (solid
stars) are taken from \cite{Afanasiev:2002mx,Appelshauser:1998yb},
and the open stars represent the data reflected at mid-rapidity. The
dash-dot-dotted lines are the cascade mode calculations, the dashed
lines represent the calculations with SM-EoS of confined baryons,
while the solid lines illustrate the results with the additional
density-dependent potentials for ``pre-formed" particles. While the
cascade calculations give a reasonable rapidity distribution for the
pions, the rapidity distribution of net protons is Gaussian - as
seen in previous UrQMD calculations \cite{Bratkovskaya:2004kv}, and
not fully consistent with the data (which has a two-bump structure
at $y\sim \pm1.2$). After considering the baryonic potentials, the
distribution only changes slightly at projectile and target
rapidities. However, when the density-dependent EoS for
``pre-formed" particles is additionally taken into account, one
observes that two bumps in the rapidity distribution of net protons
appear. This implies that, due to the interactions of ``pre-formed"
particles at early stage, the additional pressure increases the
effective stopping power as well as the central density at the
compression stage. The total pion multiplicity remains basically
unchanged on a level below $5\%$ at this beam energy. Thus, the
current pion distribution with potentials is acceptable for further
investigations.

\begin{figure*}
\includegraphics[angle=0,width=0.9\textwidth]{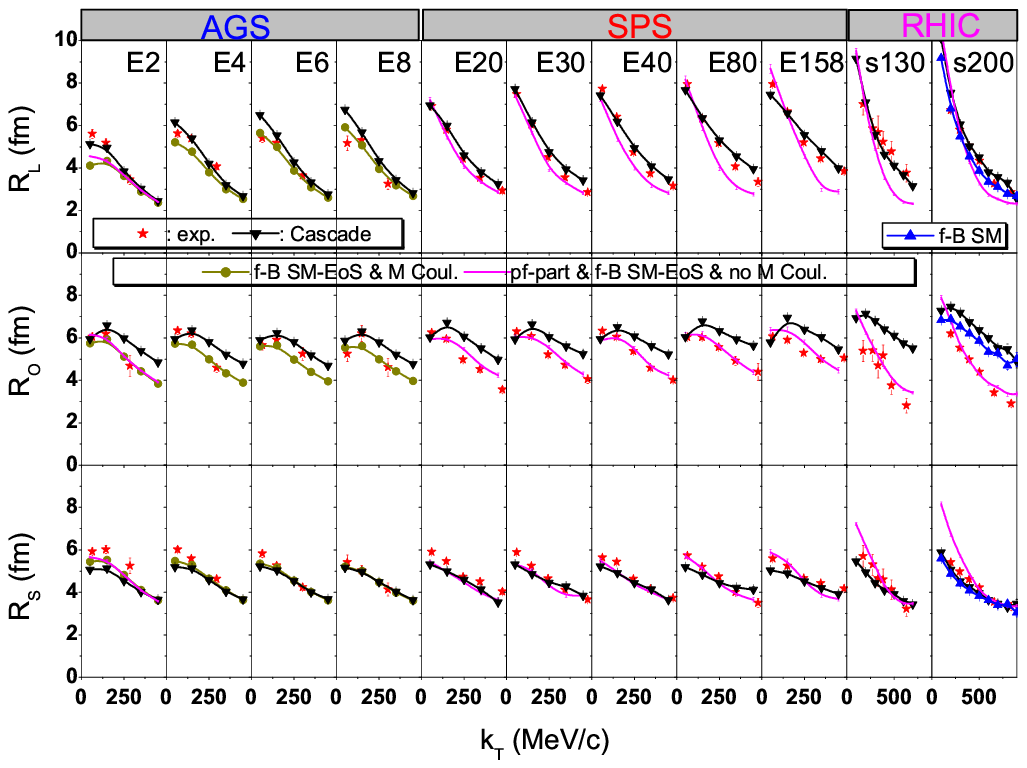}
\caption{Transverse momentum $k_T$ dependence of the HBT radii
$R_L$, $R_O$, and $R_S$ (at midrapidity) for central HICs at AGS
($E_b=2$A, $4$A, $6$A, and $8$A GeV), SPS ($E_b=20$A, $30$A, $40$A,
$80$A, and $158$A GeV), and RHIC ($\sqrt{s_{NN}}=130$ and $200$ GeV)
energies. The data are indicated by solid stars, which are from
E895, NA49, STAR, and PHENIX collaborations
\cite{Lisa:2000hw,Kniege:2006in,Adler:2001zd,Adcox:2002uc,Adler:2004rq,Adams:2004yc}.
The calculations with cascade mode are shown by lines with
down-triangles. At AGS energies, the calculations with potentials
for confined baryons (SM-EoS) as well as a Coulomb potential for
confined mesons are shown by the lines with circles. At AGS $E_b=2$A
GeV, SPS, and RHIC energies, the results with potentials for both
``pre-formed" particles and confined baryons but without Coulomb
potential for mesons are shown by lines, while the lines with
up-triangles correspond to the calculations without any potentials
for ``pre-formed" particles but with potentials for confined
baryons. } \label{fig2}
\end{figure*}

Using the analyzing program ``correlation after-burner" (CRAB
v3.0$\beta$), which is contributed by S. Pratt
\cite{Pratt:1994uf,Pratthome}, we calculate the HBT correlator and
further fit it as a three-dimensional Gaussian form under the Pratt
convention, i.e., the longitudinally comoving system (LCMS), which
is expressed as

\begin{equation}
C(q_O,q_S,q_L)=1+\lambda {\rm
exp}(-R_L^2q_L^2-R_O^2q_O^2-R_S^2q_S^2-2R_{OL}^2q_Oq_L).
\label{fit1}
\end{equation}
In Eq.~(\ref{fit1}) the $\lambda$ is the incoherence factor, $R_L$,
$R_O$, and $R_S$ are the Pratt radii in longitudinal, outward, and
sideward directions, while the cross-term $R_{OL}$ plays a role at
large rapidities. $q_i$ is the pair relative momentum in the $i$
direction. Fig.\ \ref{fig2} shows the transverse momentum $k_T$
dependence ($\textbf{k}_T=(\textbf{p}_{1T}+\textbf{p}_{2T})/2$) of
the HBT-radii $R_L$, $R_O$, and $R_S$ (at midrapidity) for central
HICs at AGS ($E_b=2$A, $4$A, $6$A, and $8$A GeV), SPS ($E_b=20$A,
$30$A, $40$A, $80$A, and $158$A GeV), and RHIC ($\sqrt{s_{NN}}=130$
and $200$ GeV) energies. The physical cuts used are same as those
listed in \cite{Li:2007im}. The data are from the E895, NA49, STAR,
and PHENIX collaborations
\cite{Lisa:2000hw,Kniege:2006in,Adler:2001zd,Adcox:2002uc,Adler:2004rq,Adams:2004yc}.
The calculations with cascade mode are shown by lines with
down-triangles while the calculations with various potential
treatments are shown with other lines with different symbols.  At
$E_b=2$A GeV, we show a comparison of calculations between with and
without Coulomb potentials for confined mesons. It is seen that the
two-body mesonic Coulomb potential before freeze-out affects the HBT
radii only very weakly at about $k_T<100$MeV$/c$, and the ratio
between $R_O$ and $R_S$ values is not altered consequently. At AGS
energies, besides the nuclear potentials, the Coulomb potential of
confined mesons is also considered (solid lines with circles), while
at SPS and RHIC energies,  the Coulomb potential for confined mesons
is switched off in order to save computing times and to avoid
problems with non-local interactions. It is known that, below AGS
energies hadrons are dominantly produced from the decay of
resonances so that the considered potentials for ``pre-formed"
particles from string fragmentations have no effect on the HBT radii
\cite{Weber:1998zb,Petersen:2006mp}. At higher energies
``pre-formed" particles play a dominant role for the early stage
dynamics, which can be seen, e.g., by the comparison of solid lines
and the lines with triangles at $\sqrt{s_{NN}}=200$ GeV. So far the
question of a steeper $k_T$-dependence of HBT radii than data at
high SPS and RHIC energies, similar to the AMPT transport model
calculations with parton scatterings \cite{Lin:2002gc}, is still
open.

The most exciting results show up in the transverse space. Roughly
speaking, the nuclear potentials lead to a smaller $R_O$ (especially
at large $k_T$) but a larger $R_S$ (especially at small $k_T$)
generally allowing for a better description of the $k_T$ dependent
data. Certainly, our present crude approach is only semi-realistic
as a description of the dynamic processes at the early stage of the
reaction. However, one can still come to the conclusion that the
potentials help to obtain a better description of the pion
freeze-out in the transverse space.

\begin{figure}
\includegraphics[angle=0,width=0.4\textwidth]{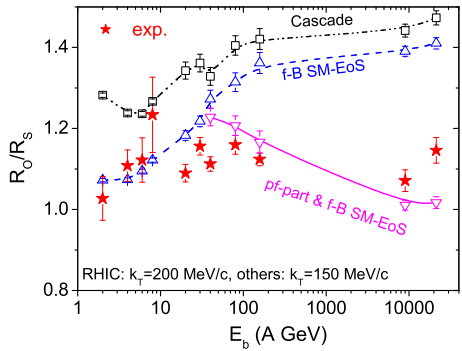}
\caption{Excitation function of the $R_O/R_S$ ratio at small $k_T$.
The data are indicated by solid stars. The dash-dot-dotted lines
with open rectangles are results under cascade mode, the dashed
lines with open up-triangles represent the calculations with
potentials for confined baryons, while the solid lines with open
down-triangles are results with potentials for both ``pre-formed"
and confined particles. } \label{fig3}
\end{figure}

In Fig.\ \ref{fig3} we depict the excitation function of the
$R_O/R_S$ ratio at small $k_T$: at AGS and SPS energies, the results
at $k_T~150\pm 50$MeV$/c$ are shown, while at RHIC energies, the
$k_T$ is set to $200\pm 50$MeV$/c$. The experimental data within the
transverse momentum regions are compared with the calculations with
and without potentials. As seen before
\cite{Li:2006gp,Li:2006gb,Li:2007im}, in the cascade mode
(dash-dot-dotted lines with open rectangles) the $R_O/R_S$ ratio,
which is related to the duration of the emission $\tau$ ($\tau \sim
\sqrt{R_O^{2}-R_S^{2}}$), is larger than the experimentally observed
values at all investigated energies. Furthermore, the increase of
the $R_O/R_S$ ratio with beam energies at AGS is not seen in these
calculations. When the SM-EoS is considered for confined baryons
(dashed lines with open up-triangles), the $R_O/R_S$ ratio decreases
compared to the result in the cascade mode and reproduces the energy
dependence of the data up to the lower SPS energies. At high SPS and
RHIC energies, however, the $R_O/R_S$ ratio increases further with
increasing beam energies and deviates strongly from the data at top
RHIC energies. Here, the ratio nearly approaches the value obtained
in the cascade mode. It implies that the potential of confined
hadrons is increasingly losing its importance with increasing beam
energy. At SPS and RHIC energies the deviation from data can be
interpreted by the absence of interactions of ``pre-formed"
particles from string fragmentation since at $E_b\sim30-40$A GeV the
measured $R_O/R_S$ ratio begins to flatten/decrease in contrast to
the aforementioned calculations. However, with the consideration of
a density dependent EoS for ``pre-formed" hadrons (solid lines with
open down-triangles) this behavior can be reproduced in the
calculations. With this approach, at RHIC energies, the ratio
$R_O/R_S$ is about unity (slightly below data) which implies that
the compressibility provided by the currently employed potential is
too strong. Nevertheless, since we are not aiming at fitting the
whole excitation function of the data, we leave this detail aside
for further detailed explorations.

Therefore, our explanation of the HBT time related puzzle is
different from the investigations in Ref.\ \cite{Cramer:2004ih}
where Cramer, et al., considered a single-pion optical potential
which simulates its interactions with the dense medium. While in
this work, the microscopic ``nuclear" potentials are considered for
``pre-formed" hadrons from the string fragmentation. The Hamilton's
equations of motion of the particles represented by Gaussian wave
packets are solved microscopically. After that, the freeze-out phase
space is analyzed numerically by the CRAB program, in which the
plain and distorted (by Coulomb and nuclear potentials after
freeze-out) wave functions are optional.

\begin{figure}
\includegraphics[angle=0,width=0.4\textwidth]{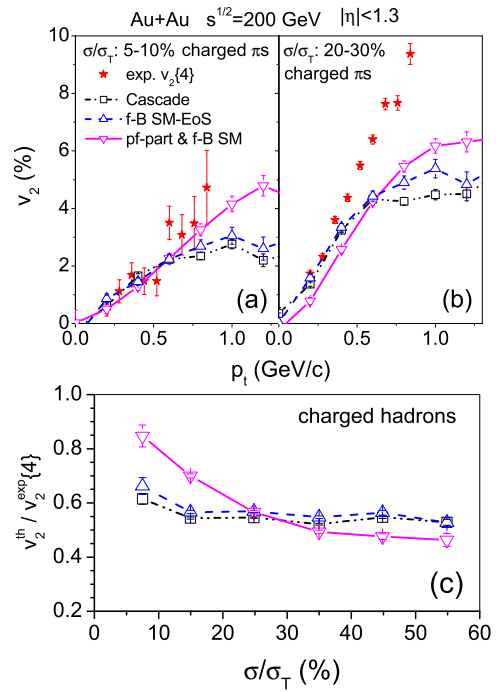}
\caption{Top plots ((a) and (b)): Transverse momentum $p_t$
dependence of the elliptic flow $v_2$ for Au+Au collisions at RHIC
$\sqrt{s_{NN}}=200$ GeV. The pseudo-rapidity cut is $|\eta|<1.3$.
Two sets of centralities are chosen: $5\%<\sigma/\sigma_T<10\%$ (in
(a)) and $20\%<\sigma/\sigma_T<30\%$ ((b)). Stars represent the
four-particle integral cumulant $v_2$ data ($v_2\{4\}$).
Dash-dot-dotted lines with open rectangles represent the cascade
calculations. Dashed lines with open up-triangles are the results
with confined baryonic potentials, while the solid lines with open
down-triangles are the results with potentials for both
``pre-formed" particles and confined baryons. Bottom plot ((c)):
Centrality dependence of the ratio between theoretical and
experimental flows of charged hadrons, which are integrated over
$p_t$ and $\eta$.} \label{fig4}
\end{figure}

Let us finally check the modification of the early transverse
expansion by a look into the elliptic flow. In the upper plots of
Fig.\ \ref{fig4} we present the transverse momentum $p_t$ dependence
of the elliptic flow $v_2$
($v_2=\langle(p_{x}^{2}-p_{y}^{2})/p_t^{2}\rangle$, where
$p_t^{2}=p_{x}^{2}+p_{y}^{2}$) of charged pions from Au+Au
collisions at RHIC $\sqrt{s_{NN}}=200$ GeV. Two sets of centralities
are chosen: $5\%<\sigma/\sigma_T<10\%$ (in (a)) and
$20\%<\sigma/\sigma_T<30\%$ ((b)). To avoid non-flow effects
\cite{Adams:2004bi}, the four-particle integral cumulant $v_2$ data
($v_2\{4\}$) are shown with stars. Similar to those in Fig.\
\ref{fig3}, the UrQMD calculations without and with potentials are
illustrated for comparison with data. It is clear that due to the
absence of early pressure, the elliptic flow in the cascade mode
(dash-dot-dotted lines with open rectangles) can not grow as fast as
the data with the increase of transverse momentum in both
centralities. After switching on the SM-EoS for confined baryons
(dashed lines with open up-triangles), the flow is seen to increase
slightly especially at high transverse momenta. With the density
dependent potentials for ``pre-formed" particles (solid lines with
open down-triangles), the elliptic flow at large transverse momenta
rises further, however, at small transverse momenta it is driven
down and deviates from data especially at large centralities. This
phenomenon calls for a more complete description of the dynamics of
HICs at high energies in which the (in)elastic scatterings of
particles (confined hadrons and deconfined partons) should be
considered in more detail. To make a whole comparison, we show in
Fig.\ \ref{fig4} (c) the centrality dependence of the ratio between
different theoretical $v_2$ values of charged hadrons and the
four-particle cumulant data (``$v_2^{th}/v_2^{exp}\{4\}$", here the
flows are integrated over $p_t$ and $\eta$). As stated before, the
flow obtained from the cascade calculation is only about $55\%$ of
the experimental one. The contribution of potentials of confined
baryons to flow is only at the level of about $4\%$. While the
potentials of ``pre-formed" hadrons increase the flow in more
central collisions. At large centrality, the potential effects are
relatively weak.

In summary, besides a soft equation of state with momentum
dependence, which is required especially at low energies, a density
dependent (Skyrme-like) potential has been tested for the
``pre-formed" particles from string fragmentation in the UrQMD
transport model. Although the form of the potential for the new
phase is simple and rough (ideally, the EoS of the new phase should
be based on the first-principle lattice QCD calculations
\cite{Bluhm:2007nu}), it provides  new insights into the dynamics of
HICs. It reproduces the proper stopping power for net protons. It
also improves the elliptic flow at large transverse momenta. Most
importantly however, it decreases the HBT radius $R_O$ but increases
the $R_S$ so that the experimental duration time related quantity
--- here the $R_O/R_S$ ratio --- can be reasonably reproduced
throughout the energies from AGS, SPS, up to RHIC. Although some
open problems remain due to the simple treatment of the interaction
of the particles at the early stage.

\section*{Acknowledgements}
We would like to thank S. Pratt for providing the CRAB program and
acknowledge support by the Frankfurt Center for Scientific Computing
(CSC). We also thank M Gyulassy, H J Drescher, and S Haussler for
helpful discussions. Q. Li thanks the Frankfurt Institute for
Advanced Studies (FIAS) for financial support. This work is partly
supported by GSI, BMBF, and Volkswagenstiftung.

\end{document}